\documentclass[prapplied,letterpaper,twocolumn,showpacs]{revtex4-1}
\usepackage{times,xspace}  
\usepackage{amsbsy,amssymb,amsmath,bm,bbold}
\usepackage{graphicx,color,epsfig,rotate} 
\usepackage{fancyhdr} 
\usepackage{epstopdf}
\usepackage{csquotes}
\usepackage{float}
\usepackage[colorlinks=true,linkcolor=blue,citecolor=blue,urlcolor=blue]{hyperref}

\def\bbbc{{\mathchoice {\setbox0=\hbox{$\displaystyle\rm C$}\hbox{\hbox 
				to0pt{\kern0.4\wd0\vrule height0.9\ht0\hss}\box0}} 
		{\setbox0=\hbox{$\textstyle\rm C$}\hbox{\hbox 
				to0pt{\kern0.4\wd0\vrule height0.9\ht0\hss}\box0}} 
		{\setbox0=\hbox{$\scriptstyle\rm C$}\hbox{\hbox 
				to0pt{\kern0.4\wd0\vrule height0.9\ht0\hss}\box0}} 
		{\setbox0=\hbox{$\scriptscriptstyle\rm C$}\hbox{\hbox 
				to0pt{\kern0.4\wd0\vrule height0.9\ht0\hss}\box0}}}}

\begin{document}
	
\title{Third-order Exceptional Point and Successive Switching among Three States in a Degenerate Optical Microcavity}
\author{Arnab Laha,$^{1,3}$ Dinesh Beniwal,$^2$ Sibnath Dey,$^1$ Abhijit Biswas,$^3$ and Somnath Ghosh$^{1,}$}
\email{somiit@rediffmail.com}
\affiliation{\vspace{0.3cm}\\$^1$Department of Physics, Indian Institute of Technology Jodhpur, Rajasthan-342037, India\\$^2$Department of Physics, National Institute of Science Education and Research Bhubaneswar, Odisha 752050, India\\$^3$Institute of Radiophysics and Electronics, University of Calcutta, Kolkata-700009, India}

\begin{abstract}
	One of the most intriguing topological features of open systems is exhibiting exceptional point (EP) singularities. Apart from the widely explored second-order EPs (EP2s), the explorations of higher-order EPs in any system requires more complex topology, which is still a challenge. Here, we encounter a third-order EP (EP3) for the first time in a simple fabrication feasible gain-loss assisted optical microcavity. Using scattering-matrix formalism, we study the simultaneous interactions between three successive coupled states around two EP2s, which yield an EP3. Following an adiabatic parametric variation around the identified EP3, we present a robust successive-state-conversion mechanism among three coupled states. The proposed scheme indeed opens a unique platform to manipulate light in integrated devices.   
\end{abstract}

\maketitle

\section{Introduction}

Exceptional points (EPs) are the branch point singularities that appear as topological defects in parameter space of various open or non-Hermitian systems \cite{Kato95,Heiss00,Heiss12,Rotter15,Ali19,Ozdemir19,Gupta19}. While passing through an EP, coupled eigenvalues and the corresponding eigenvectors of such a system simultaneously coalesce; which makes the underlying Hamiltonian defective \cite{Kato95,Heiss00,Heiss12,Rotter15,Ali19,Ozdemir19,Gupta19}. A stroboscopic parametric variation enclosing a second-order EP (EP2) significantly affects the dynamics of the corresponding pair of coupled states, where one complete parametric encirclement allows the adiabatic permutation between them with acquisition of additional Berry phase \cite{Dembowski04}. Now, during such encirclement, if we consider dynamical (i.e., time or length-scale dependent) parametric variation around an EP2, then the system behavior which follows the adiabaticity otherwise, breaks down yielding the chiral dynamics with time-asymmetric population transfer between the coupled states \cite{Gilary13}. Such unique topological features of EPs have been exploited in-depth to meet a wide range of technological challenges like,  asymmetric-mode-conversion \cite{Doppler16,ZhangX18,Gandhi20,Ghosh16,Laha18}, topological state-switching \cite{Laha17,Laha17a,Laha19}, lasing-control \cite{Hodaei15}, unidirectional propagation with enhanced nonreciprocity \cite{Thomas16,Laha20}, sensitivity enhancement \cite{Wiersig16,Chen17,Hokmabadi19}, etc.

Instead of only EP2s, intense theoretical efforts have been put forward to investigate higher-order EPs \cite{Demange12,Menke16,Ryu12,Bhattacherjee19_1,Bhattacherjee19_2,Bhattacherjee19_3}. Here, an EP of order $n$ (EP$n$) is defined as the simultaneous coalescence of $n$ coupled states, however, on the other way, with the presence of $(n-1)$ EP2s, the topological effects of an EP$n$ can also be realized \cite{Menke16,Ryu12,Bhattacherjee19_1,Bhattacherjee19_2,Bhattacherjee19_3}. In this context, third-order EPs (EP3s) have garnered attention, as the cube-root response near an EP3 is extremely sensitive to the external perturbation, in comparison to the square-root response near an EP2 \cite{Hodaei17,Zeng19}. The integration of an EP3 in any real system and realization of successive conversions among three states due to parametric encirclement require a complicated system topology that have been shown using individually pumped coupled-cavity \cite{Jing17} and waveguide \cite{Heiss16,Schnabel17} systems. Recently, the nonadiabatic state-dynamics due to dynamical encirclement around an EP3 has also been explored in coupled \cite{ZhangX19} and planar \cite{Dey20} waveguide systems. However, there is still a technological demand to locate an EP3 with associated successive state conversion in a simple, fabrication feasible system having minimum number of control parameters.    

In this letter, we report the hosting of an EP3 in a gain-loss assisted Fabry-P\'erot type trilayer optical microcavity. With proper distribution and simultaneous variation of gain-loss profile, the mutual interaction between three chosen cavity-states has been studied around two EP2s. Here, the effect of an EP3 has been realized with simultaneous presence of these two EP2s inside a closed gain-loss parameter space. Beyond the reported systems with more complex topology \cite{Jing17,Heiss16,Schnabel17,ZhangX19}, we present a robust successive topological conversion between three consecutive states around an EP3, for the first time, using only two tunable parameters associated with the imposed gain-loss profile. Proposed design hosting an EP3 having rich physics and unique features will certainly be suitable for state-of-the-art device implementation.

\section{Analytic structure of an EP3 in presence of multiple EP2s}

To establish the condition of the presence of an EP3 with simultaneous presence of two EP2s, we consider a three-level open system characterized by a $3\times3$ non-Hermitian Hamiltonian $\mathcal{H}$ as
\begin{equation}
\mathcal{H}=\left(\begin{array}{ccc}\varepsilon_1 & 0 & 0 \\0 & \varepsilon_2 & 0 \\0 & 0 & \varepsilon_3\end {array}\right)+\lambda\left(\begin{array}{ccc}0 & -\eta_p  & 0 \\ \delta\eta_p & 0 & \delta\eta_q \\ 0 & -\eta_q & 0\end {array}\right).
\label{eq_H}
\end{equation}   
Here $\mathcal{H}$ represents a passive system, consisting of three passive eigenvalues $\varepsilon_j\,(j=1,2,3)$, which is influenced by a complex perturbation. In the perturbation element, $\eta_p$ and $\eta_q$ are two complex coupling terms that are connected through a tunable parameter $\delta$. $\lambda$ represent the perturbation strength. With proper optimization of these parameters, the perturbation can be analogous to the optical gain-loss. Now the eigenvalues of $\mathcal{H}$ can be calculated from the following cubic equation given by 
\begin{equation}
E^3+m_1E^2+m_2E+m_3=0;
\label{eq_E}
\end{equation}
with 
\begin{subequations}
\begin{align}
m_1&=-\left(\varepsilon_1+\varepsilon_2+\varepsilon_3\right),
\label{eq_m1}\\
m_2&=\varepsilon_1\varepsilon_2+\varepsilon_2\varepsilon_3+\varepsilon_3\varepsilon_1+\lambda^2\delta\left(\eta_p^2+\eta_q^2\right),
\label{eq_m2}\\
m_3&=-\varepsilon_1\varepsilon_2\varepsilon_3-\lambda^2\delta\left(\eta_p^2\varepsilon_3+\eta_q^2\varepsilon_1\right).
\label{eq_m3}
\end{align}
\end{subequations}
Now, three roots of Eq. \ref{eq_E}, say $E_j\,(j=1,2,3)$, give three eigenvalues of $\mathcal{H}$ (given by Eq. \ref{eq_H}). Using Cardano's method~\cite{Korn}, we can calculate $E_j\,(j=1,2,3)$ as
\begin{subequations}
	\begin{align}
	E_1&=\omega\kappa_++\bar{\omega}\kappa_--\xi,
	\label{eq_Eig1}\\
	E_2&=\kappa_++\kappa_--\xi,
	\label{eq_Eig2}\\
	E_3&=\bar{\omega}\kappa_++\omega\kappa_--\xi.
	\label{eq_Eig3}
	\end{align}
	\label{eq_Eig}
\end{subequations} 
$\omega$ is the cube root of unity: $\omega^3=1$. $\bar\omega$ represents the complex conjugate of $\omega$. Here,
\begin{equation}
\kappa_{\pm}=\left(u\pm\sqrt{u^2+v^3}\right)^{1/3}\quad\textnormal{and}\quad\xi=m_1/3;
\label{eq_kappa-xi} 
\end{equation} 
with
\begin{equation}
u=-\frac{m_1^2}{27}+\frac{m_1m_2}{6}-\frac{m_3}{6}\quad\textnormal{and}\quad v=-\frac{m_1^2}{9}+\frac{m_2}{3}. 
\end{equation}     
Now among three coupled eigenvalues $E_j\,(j=1,2,3)$, we can individually control the interaction between the pairs $\{E_1,\,E_2\}$, as well as $\{E_2,\,E_3\}$. In the Hamiltonian $\mathcal{H}$ (given by Eq. \ref{eq_H}), we deliberately neglect the interaction between $E_1$ and $E_3$. Here, two different EP2s can be identified with the coalescence of $E_2$ with either $E_1$, or $E_3$, individually for two different settings of perturbation. Such situations entitle the validity of the conditions given by
\begin{equation}
\kappa_+=\kappa_-\quad\textnormal{and}\quad\omega\kappa_+=\kappa_-\,\,\textnormal{or}\,\,\bar{\omega}\kappa_+=\kappa_-.
\label{eq_EP}
\end{equation}      
Now, the validity of the equalities given in Eq.~\ref{eq_EP} implies that the square root part of $\kappa_{\pm}$ in Eq. \ref{eq_kappa-xi} is equal to zero. Under such conditions of the hosting of two different EP2s, the cube root nature of $\kappa_{\pm}$ indicate the presence of an analogous cube root branch point, i.e., an EP3, at which the coupled eigenvalues (given by Eq. \ref{eq_Eig}) are analytically connected. In the following section, we implement the above {\it scenario} in a simple fabrication feasible partially-pumped optical microcavity. 

\section{Results and discussion}

\subsection{Design of a Fabry-P\'erot type optical microcavity with scattering matrix formalism}

Here, we present a Fabry-P\'erot type 1D two-port open trilayer optical microcavity that has been partially pumped by the non-Hermiticity introduced in terms of an unbalanced gain-loss profile (beyond PT-symmetry). A high-indexed material ($n_s$) has been sandwiched between two same low indexed layers ($n_g$) to realize our trilayer cavity system of the width $L$. Here, we choose $n_s=3.48$, $n_g=1.5$ and $L=12\,\mu m$. The unbalanced gain-loss profile has been introduced in two low indexed layers, where the width of each layer is $5.05\,\mu m$. There is no gain-loss in the intermediate high indexed layer of width $1.9\,\mu m$.
\begin{figure}[b!]
	\centering
	\includegraphics[width=8.6cm]{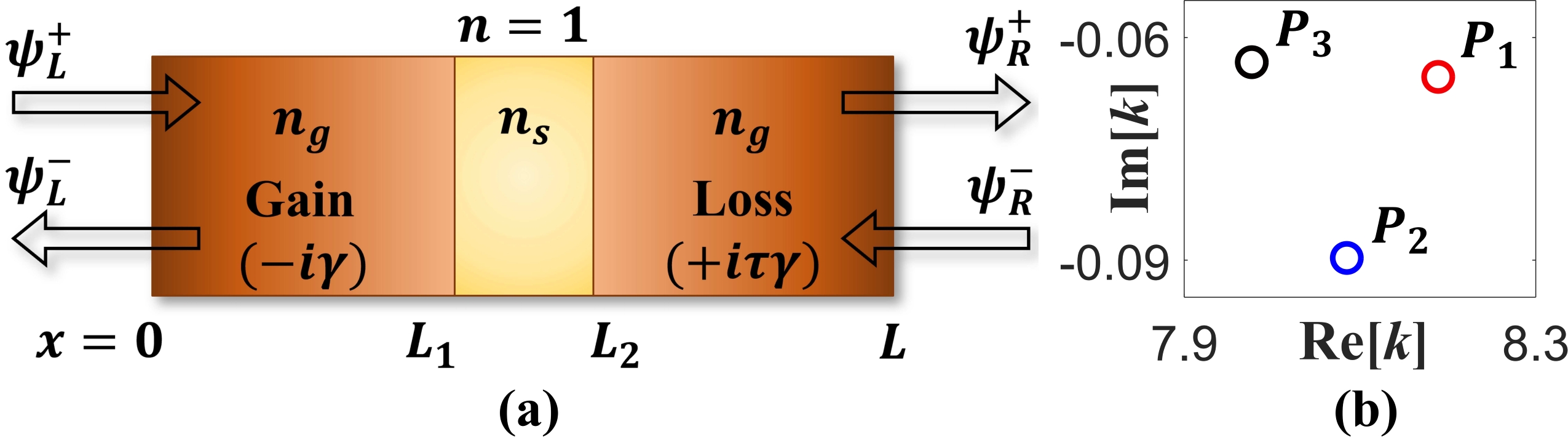}
	\caption{\textbf{(a)} Schematic of the proposed microcavity with $n_g=1.5$, $n_s=3.48$, $L=12\,\mu m$, $L_1=5.05\,\mu m$ and $L_2=6.95\,\mu m$. $\left\{\psi_{L}^+,\psi_{R}^-\right\}$ and $\left\{\psi_{L}^-,\psi_{R}^+\right\}$ represent the sets of incident and scattered waves, respectively from both sides. \textbf{(b)} Initial positions of three coupled poles, say, $P_1$, $P_2$ and $P_3$ within the chosen frequency range from 7.9 to 8.3 $\mu m^{-1}$.}
	\label{fig:1}
\end{figure}
To characterize the imposed gain-loss profile, we introduce two tunable parameters, namely, gain-coefficient ($\gamma$) and fractional loss-to-gain ratio ($\tau$). Now, the overall nonuniform distribution of real-refractive index with associated gain-loss profile in the designed microcavity, occupying the region $0\le x\le L$, can be represented as   
\begin{equation}
n(x)=\left\{ 
\begin{array}{ll}
n_g-i\gamma &\quad\text{for}\,\,\, 0\le x\le L_1,\\
n_s &\quad\text{for}\,\,\, L_1\le x\le L_2,\\
n_g+i\tau\gamma &\quad\text{for}\,\,\, L_2\le x\le L
\end{array}
\right.
\label{nx} 
\end{equation}
with $L_1=5.05\,\mu m$ and $L_2=6.95\,\mu m$. The independent modulation of imaginary part of the refractive index irrespective of the choices of fixed passive refractive indices is dictated by Kramers-Kronig causality condition only at a single frequency \cite{Phang15}. With appropriate scalability, a similar prototype can be fabricated by a combination of silica-based glass material with pure silicon, where the customized gain-loss profile can be furnished using standard photolithography technique or by controlled doping of gain/lossy elements.

To calculate the cavity-resonances, we employ scattering-matrix ($S$-matrix) formalism method \cite{Bykov12,Laha16} and accordingly, we construct a $2\times 2$ $S$-matrix associated with the designed cavity. Considering, $\left\{\psi_{L0}^+,\psi_{R0}^-\right\}$ and $\left\{\psi_{L0}^-,\psi_{R0}^+\right\}$ as the amplitudes of corresponding incident and scattered waves form both the sides (as shown in Fig. \ref{fig:1}), the input-output relation through the $S$-matrix can be written as  
\begin{equation}
\left(\begin{array}{c}\psi_{L0}^-\\\psi_{R0}^+\end {array}\right)=S(n(x),k)\left(\begin{array}{c}\psi_{L0}^+\\\psi_{R0}^-\end {array}\right);
\label{S}
\end{equation}
Here, the matrix elements have been calculated in terms of refractive index (given by Eq. \ref{nx}) and frequency ($k$), using the electromagnetic scattering theorem. According to the $S$-matrix formalism method, the complex poles, appeared in fourth quadrant of the complex $k$-plane, represent the physical cavity resonances. Using numerical root finding method, we have calculated such matrix-poles by solving the equation $1/\left[\max|\text{eig}(S)|\right]=0$. 

Now to study the three-state interaction phenomena within the designed cavity, we have judiciously chosen three specific poles within $k$-range from 7.9 to 8.3 $\mu m^{-1}$. Initial locations of these poles, namely $P_1$, $P_2$ and $P_3$ in complex $k$-plane have been shown in Fig. \ref{fig:1}(b). Now, all these poles are mutually coupled with the onset of gain-loss. To initiate such mutual coupling between three consecutive states with gain-loss variation, we have to ensure the nonlinear initial distribution of corresponding $S$-matrix poles in complex $k$-plane as shown in Fig. \ref{fig:1}(b), which is entirely dependent on the choice of nonuniform real refractive index profile \cite{Bykov12}.

\subsection{Encounter of multiple EP2s}

\begin{figure}[t]
	\centering
	\includegraphics[width=8.6cm]{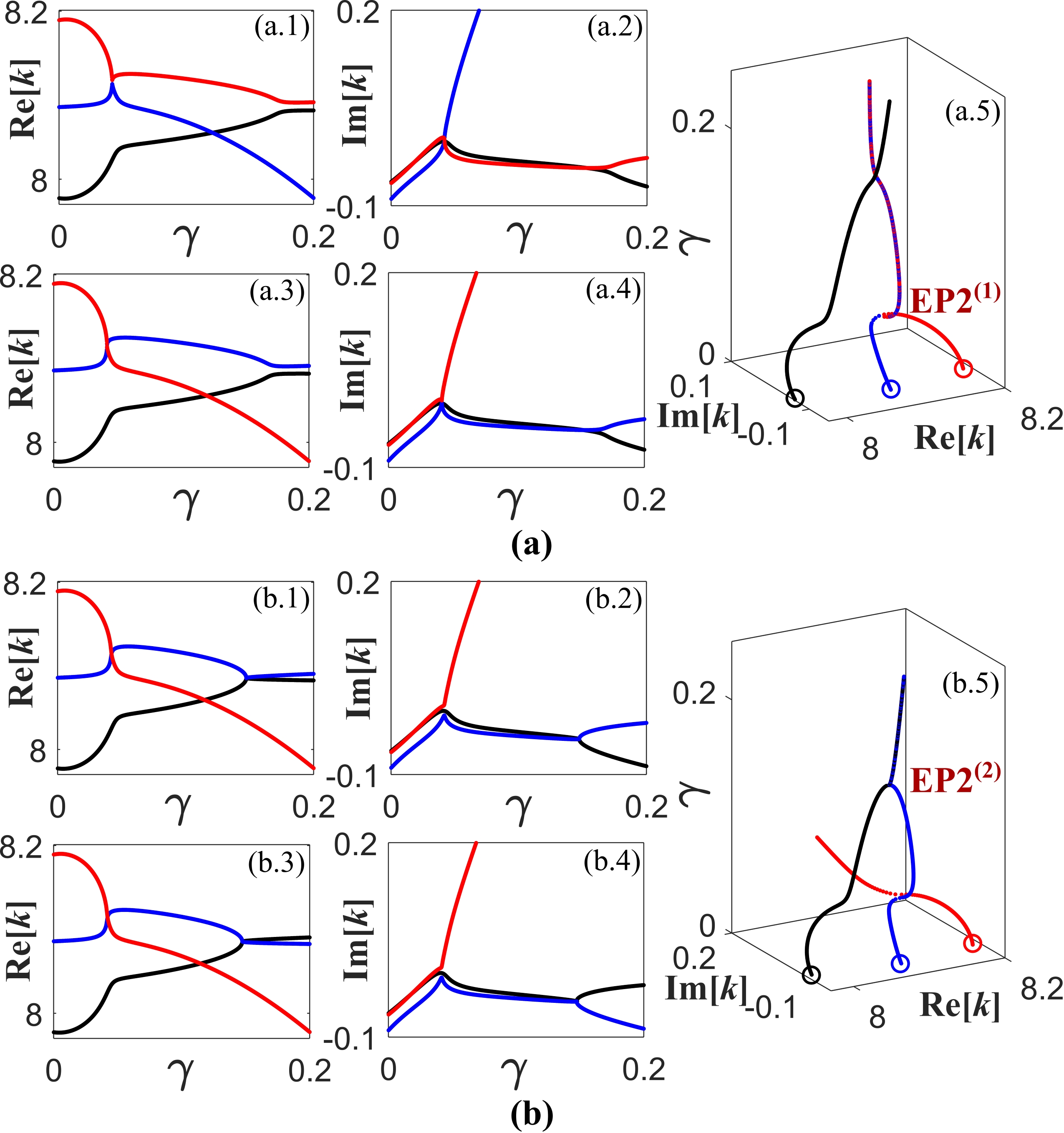}
	\caption{Trajectories of the complex $k$-values of $P_j\,(j=1,2,3)$ shown by red, blue and black curves, respectively. \textbf{(a)} $P_1$ and $P_2$ interact strongly unaffecting $P_3$. During interactions, $P_1$ and $P_2$ exhibit simultaneous (a.1) anticrossing in Re[$k$] and (a.2) crossing in Im[$k$] for $\tau=0.05$, whereas, simultaneous (a.3) crossing in Re[$k$] and (a.4) anticrossing in Im[$k$] for $\tau=0.051$. (a.5) For $\tau=0.0505$, $P_1$ and $P_2$ coalesce at $\gamma\approx0.0418$ keeping $P_3$ as an observer, that indicate the location of EP2$^{(1)}$ between $P_1$ and $P_2$. \textbf{(b)} Unaffecting $P_1$, $P_2$ and $P_3$ interact strongly. During interactions, $P_2$ and $P_3$ exhibit simultaneous (b.1) anticrossing in Re[$k$] and (b.2) crossing in Im[$k$] for $\tau=0.0585$, whereas, simultaneous (b.3) crossing in Re[$k$] and (b.4) anticrossing in Im[$k$] for $\tau=0.0591$. (b.5) For $\tau=0.0588$, $P_2$ and $P_3$ coalesce at $\gamma\approx0.1475$ keeping $P_1$ as an observer, that indicate the location of EP2$^{(2)}$ between $P_2$ and $P_3$.}
	\label{fig:2}
\end{figure}

In Fig. \ref{fig:2}, we investigate the interactions among three coupled poles by tracking their trajectories in $k$-plane for different $\tau$-values, while varying $\gamma$ in between [0, 0.25]. Red, blue and black curves indicate the trajectories of $P_1$, $P_2$ and $P_3$, respectively. In Fig. \ref{fig:2}(a), we have shown that $P_1$ and $P_2$ exhibit two topologically different special avoided resonance crossings (ARCs) in complex $k$-plane for two different $\tau$-values, where $P_3$ remains away from these ARC regimes. For $\tau=0.05$, $P_1$ and $P_2$ experience an anticrossing in Re[$k$] as shown in plot \ref{fig:2}(a.1) with a simultaneous crossing in Im[$k$] as shown in plot \ref{fig:2}(a.2). Now with a slightly increasing $\tau$ to 0.051, we can observe an exactly opposite topological behavior depicting a crossing in Re[$k$] and an simultaneous anticrossing in Im[$k$] of $P_1$ and $P_2$ in the plots (a.3) and (a.4), respectively. Such two topologically different ARCs definitely confirm the presence of an EP2, say EP2$^{(1)}$, as shown in plot \ref{fig:2}(a.5), where for an intermediate $\tau=0.0505$, $P_1$ and $P_2$ coalesce in complex $k$-plane (unaffecting $P_3$) near $\gamma\approx0.0418$. Thus, we find the location of EP2$^{(1)}$ in $(\gamma,\tau)$ plane at $\sim(0.0418,0.0505)$.

In a similar way, in Fig. \ref{fig:2}(b), we have shown two topologically dissimilar ARCs between $P_2$ and $P_3$ in complex $k$-plane for two specifically chosen $\tau$-values, for which $P_1$ remains away from these ARC regimes. During the interactions between $P_2$ and $P_3$, an anticrossing in Re[$k$] with a simultaneous crossing in Im[$k$] for $\tau=0.0585$ (as can be seen in plots \ref{fig:2}(b.1) and (b.2), respectively) and a crossing in Re[$k$] with a simultaneous anticrossing in Im[$k$] for $\tau=0.0591$ (as can be seen in plots \ref{fig:2}(b.3) and (b.4), respectively) certainly indicate the presence of another EP2, say EP2$^{(2)}$. In plot \ref{fig:2}(b.5), the location of EP2$^{(2)}$ have been found in $(\gamma,\tau)$-plane at $\sim(0.1475,0.0588)$ by pointing out the coalescence between $P_2$ and $P_3$ in complex $k$-plane (unaffecting $P_1$) near $\gamma\approx0.1475$ for $\tau=0.0588$. Thus, only tuning two control parameters $\gamma$ and $\tau$, we find two cases, where two different pairs from three coupled poles interact around two EP2s keeping the third one as an observer. Within a chosen interaction regime, this situation indicate the presence of an EP3 where all three interacting poles are analytically connected.
\begin{figure}[b!]
	\centering
	\includegraphics[width=8.6cm]{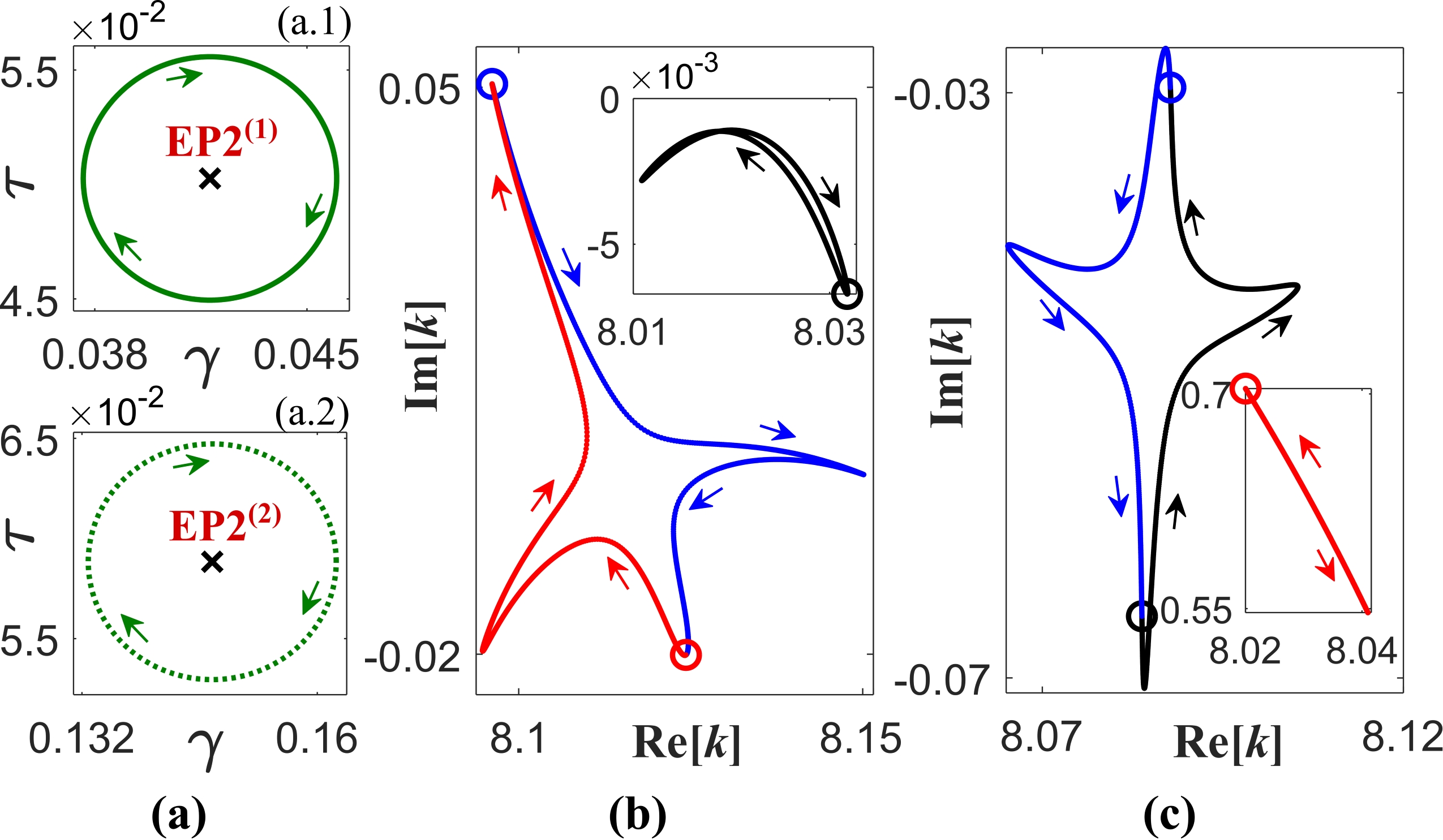}
	\caption{\textbf{(a)} Individual parametric encirclements in $(\gamma,\tau)$-plane following Eq. \ref{circle} (a.1) around only EP2$^{(1)}$ keeping EP2$^{(2)}$ outside (solid green circle) and (a.2) around only EP2$^{(2)}$ keeping EP2$^{(1)}$ outside (dotted green circle). \textbf{(b)} Dynamics of the complex $k$-values of $P_1$ (red curve), $P_2$ (blue curve) and $P_3$ (black curve; in the inset) following the encirclement process shown in (a.1): $P_1$ and $P_2$ permutes, and $P_3$ makes individual loop. \textbf{(c)} Similar dynamics of the complex $k$-values of $P_1$ (in the inset), $P_2$ and $P_3$ following the encirclement process shown in (a.2): $P_2$ and $P_3$ permutes, and $P_1$ makes individual loop. The dynamics of $P_3$ in (b) and $P_1$ in (c) have been shown in respective insets for clear visualization. In (b) and (c), red, blue and black circular markers indicate the initial positions of $P_1$, $P_2$ and $P_3$ while initializing the encirclement process. Arrows of different colors indicate the respective direction of progressions.}
	\label{fig:3}
\end{figure}  

Now, to study the branch-point behaviors of identified EP2s, we consider adiabatic encirclements around each of the EP2s individually and also simultaneously around both the EP2s in system parameter-space. The parametric equations given by 
\begin{subequations}
\begin{align}
\gamma(\theta)&=\gamma_{c}\left[1+r\cos(\theta)\right]\\
\tau(\theta)&=\tau_{c}\left[1+r\sin(\theta)\right]
\end{align}
\label{circle}
\end{subequations} 
has been considered to carry out a specific encirclement process in $(\gamma,\tau)$-plane. Here ($\gamma_c,\tau_c$) represent the center, and simultaneous stroboscopic gain-loss variation has been achieved by two characteristics parameters $r$ and $\theta$ (given that $0\le\theta\le2\pi$). With judicious choices of ($\gamma_c,\tau_c$) and $r$, we can encircle individual EP2s or both the EP2s. 

\subsection{Effect of parametric encirclement around individual EP2s}   

In Fig. \ref{fig:3}, we individually encircle EP2$^{(1)}$ and EP2$^{(2)}$ in $(\gamma,\tau)$-plane following Eq. \ref{circle} and study the dynamics of three coupled poles $P_j\,(j=1,2,3)$ in complex $k$-plane. Such two separate encirclement processes have been shown in plots (a.1) and (a.2) of Fig. \ref{fig:3}(a) by choosing centers at respective EP2s (as shown by the black crosses) and $r=0.1$ for both cases. The green arrows represent the direction of encirclements. In Figs. \ref{fig:3}(b) and (c), the evolutions of $P_1$, $P_2$ and $P_3$ have been represented by red, blue and black dotted curves in complex $k$-plane, where for a clear understanding of the state-dynamics, we indicate the initial positions of these coupled poles (while initializing the encirclement process) by circular markers and also represent their evolutions through arrows with the similar variant in colors. For a particular encirclement process, each point in the dynamics of three coupled poles in $k$-plane has been generated by each point evolution along the loop in $(\gamma,\tau)$-plane. 

In Fig. \ref{fig:3}(b), we track the dynamics of three coupled poles following the encirclement process shown in plot \ref{fig:3}(a.1) that encloses only EP2$^{(1)}$, keeping EP2$^{(2)}$ outside. As can be shown in Fig. \ref{fig:3}(b), only the poles $P_1$ and $P_2$ associated with EP2$^{(1)}$ permute in complex $k$-plane in the sense that they exchange their initial locations with one complete encirclement process around only EP2$^{(1)}$. However, such closed parametric variation shown in plot \ref{fig:3}(a.1) does not affect the dynamics of $P_3$ that remains in same state, making an individual loop in $k$-plane (as shown in the inset of Fig. \ref{fig:3}(b); for clear visualization), at the end of the encirclement process. In Fig. \ref{fig:3}(c), the trajectories of three coupled poles have been plotted following the encirclement process shown in plot \ref{fig:3}(a.2) that encloses only EP2$^{(2)}$, keeping EP2$^{(1)}$ outside. Here, we have shown that for one complete cycle following the parametric loop, the poles $P_2$ and $P_3$ that are associated with EP2$^{(2)}$ swap their initial positions, however $P_1$ makes an individual loop (as shown in the inset of Fig. \ref{fig:3}(c); for clear visualization) and remains unaffected from this encirclement process. Thus the individual encirclement around each of the EP2s in $(\gamma,\tau)$-plane allows the flipping between two corresponding coupled states in complex $k$-plane even in the presence of a nearby state, which firmly assures their second-order branch point behaviors of both the EP2s.  

\subsection{Effect of simultaneous parametric encirclement around two EP2s: Toward successive state conversion} 
\begin{figure*}[t]
	\centering
	\includegraphics[width=18.2cm]{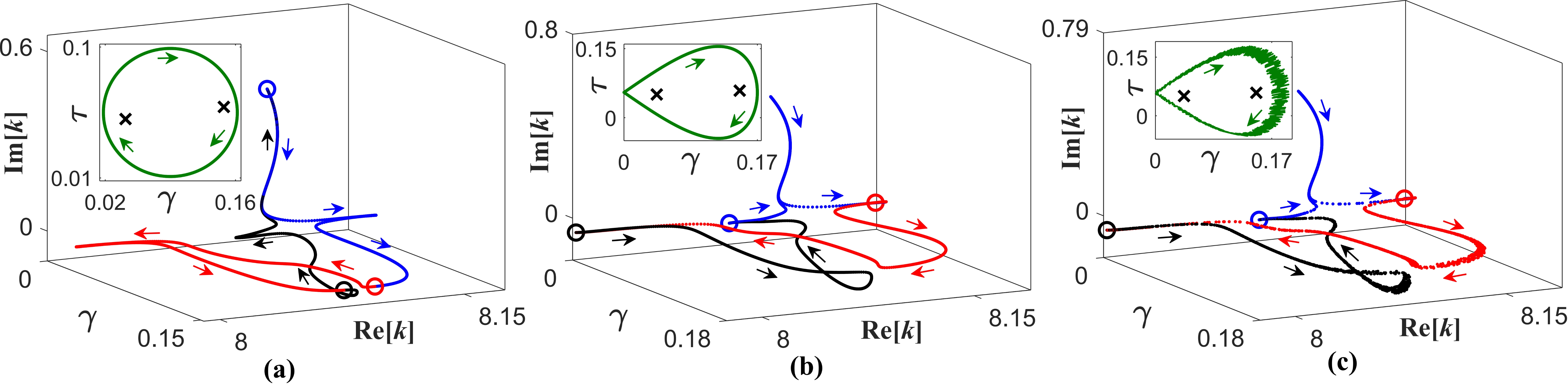}
	\caption{Dynamics of the complex $k$-values of $P_j\,(j=1,2,3\text{; shown by red, blue and black curves, respecively})$ with simultaneous parametric encirclement processes around both the EP2s following \textbf{(a)} Eq. \ref{circle}, \textbf{(b)} Eq. \ref{ellipse} and \textbf{(c)} Eq. \ref{ellipse} with an additional $5\%$ random fluctuation. For all the encirclement processes, as shown in the respective insets, $P_1$, $P_3$ and $P_2$ switch successively among them following the manner $P_1\rightarrow P_3\rightarrow P_2\rightarrow P_1$. All the colors, markers and notations carry the  same meaning as given in the caption of Fig. \ref{fig:3}.}
	\label{fig:4}
\end{figure*} 

Now, we consider the quasi-static parametric variation enclosing both the EP2s in $(\gamma,\tau)$-plane to explore the third-order branch point behavior and trace the topological dynamics of thee coupled poles in complex $k$-plane. Corresponding results have been shown in Fig. \ref{fig:4}, where the colors and notations carry the same meaning as we have already discussed in previous. In Fig. \ref{fig:4}(a), we exhibit the state-dynamics following sufficiently slow parametric evolution around both EP2$^{(1)}$ and EP2$^{(2)}$ (as shown in the inset) along the circular loop given by Eq. \ref{circle} with center at ($\gamma_c=0.09,\,\tau_c=0.055$) and $r=0.7$. Here we have shown that following one complete loop in $(\gamma,\tau)$-plane, all three coupled poles $P_1$, $P_2$ and $P_3$ exchange their identities. They switch successively following the manner $P_1\rightarrow P_3\rightarrow P_2\rightarrow P_1$ and make a complete loop in complex $k$-plane. In Fig. \ref{fig:4}(a), we also plot the variation of $\gamma$ (along an additional axis) in addition with the complex $k$-plane to show such successive state-flipping phenomenon. Such unconventional dynamics of three coupled poles around their two EP2s certainly confirm the presence of an EP3 in system parameter plane where all three chosen poles are analytically connected. The effect of parametric encirclement, as shown in Fig. \ref{fig:4}(a), reveals the third-oder branch point behavior of an EP3 because three consecutive coupled states regain their initial locations after completing three-fold parametric encirclement process in a row. 
    
Now, to explore the topological nature of the third-oder branch point behavior of the embedded EP3, we consider a different shape of closed parametric-loop to encircle both the identified EP2s in $(\gamma,\tau)$-plane and study the dynamics of three coupled poles. Accordingly we consider a parameter space following the equations given by   
\begin{equation}
\gamma(\theta)=\gamma_0\sin\left(\theta/2\right)\,\,\,\text{and}\,\,\,\tau(\theta)=\tau_{0}+r\sin(\theta).
\label{ellipse}
\end{equation}
With judicious choices of $\gamma_0$ (must be greater than the $\gamma$-value associated with EP2$^{(2)}$), $\tau_0$ and $r$, we can encircle both the EP2s in $(\gamma,\tau)$-plane. Unlike circular parameter space given by Eq. \ref{circle}, Eq. \ref{ellipse} can consider the situation $\gamma=0$ for both $\theta=0$ and $2\pi$ so that we can observe the switchings from the passive locations of the chosen poles. Thus for device implementation of state-switching phenomena using the framework of our proposed waveguide the parameter space given by Eq. \ref{ellipse} should be more convenient. Now, choosing $\gamma_0=0.17$, $\tau_0=0.055$ and $r=0.7$, we perform such an encirclement process and study the dynamics of the chosen poles in Fig. \ref{fig:4}(b). Here, we have also observed the successive switchings $P_1\rightarrow P_3\rightarrow P_2\rightarrow P_1$ in complex $k$-plane like a generic fashion as we have already shown in Fig. \ref{fig:4}(a), however, most importantly, the coupled poles switch from their passive locations for this encirclement process. Thus, the successive state-switching around an EP3 in presence of two associated EP2s is omnipresent irrespective of the shape of the parameter-space. For such a complete encirclement process following Eq. \ref{ellipse}, we can also calculate the conversion efficiencies in terms overlap integrals between the passive eigenvectors corresponding to the chosen coupled poles \cite{Laha19}. For any conversion $P_i\rightarrow P_j$, the conversion efficiency $C_{ij}$ can be defined as 
\begin{equation}
C_{ij}=\frac{\left|\int\Phi_i\Phi_j dx\right|^2 }{\int|\Phi_i|^2 dx \int|\Phi_j|^2 dx}\text{;}\quad \{i,j\}\in \{1,2,3\}\text{,}\,i\ne j.
\label{C}
\end{equation} 
Here, $\Phi_i$ and $\Phi_j$ have been considered as the eigenvectors associated with $P_i$ and $P_j$. Using Eq. \ref{C}, we calculate the conversion efficiencies for the successive conversions as shown in Fig. \ref{fig:4}(b) as $C_{13}\approx91.8\%$, $C_{32}\approx94.3\%$ and $C_{21}\approx97.1\%$.

During device implementation of the proposed EP3-aided successive state conversion scheme with state-of-the-art fabrication techniques, owing to the tolerances, the parametric variations around the closed loop of EP2s may alter. So, we deliberately incorporate random fluctuations up to $\sim5\%$ on the same parametric contour already shown in the inset of Fig. \ref{fig:4}(b) to take into account such fabrication tolerances. In Fig. \ref{fig:4}(c), we check the robustness of the successive state conversion phenomena against this parametric fluctuation during encirclement process, where the modified parameter space has been shown in the inset. Here, we also observe the successive conversions among the coupled poles in a similar manner as we have already seen in Figs. \ref{fig:4}(a) and (b); which indicates the immutable third-order branch point behavior even in presence of sufficient fabrication tolerances. Such topological behavior will be robust until the parametric fluctuations does not affect the embedded EP3. 
\vfill
\section{Summary}                

In summary, we have reported the hosting of an of third-order (EP3) in a fabrication feasible two-port open trilayer Fabry-P\'erot type optical microcavity. The interaction phenomena among three coupled cavity-states within a chosen frequency range have been modulated with an unbalanced gain-loss profile characterized by only two tunable parameters. Without using of any complex topology with many tunable parameters, we have directed a particular state to interact with two other states via two EP2s by controlling only 2D gain-loss parameter space. The second-order branch point behaviors of the identified EP2s have been verified with flip-of-states phenomena between the corresponding coupled states followed by gain-loss variations enclosing each of the EP2s individually. Now simultaneously encircling both the EP2s in the gain-loss plane, we have achieved the third-order branch point behavior of an EP3, where three coupled states switch successively among them. Using the different shapes of the parametric loops, we have established such exclusive EP3-aided robust topological state-switching phenomena among three coupled states. Our proposed cavity configuration with an EP3 will certainly open-up an advanced platform to fabricate high-performance integrated optical devices to study the singularity assisted unconventional physical effects. Moreover, the proposed scheme can also be implemented in any guided wave structures to realize mode-converters, circulators (in presence of nonlinearities), etc.

\section*{Acknowledgment}    
  
A.L. and S.G. acknowledge the financial support from the Science and Engineering research Board (SERB) [Grant No. ECR/2017/000491], Department of Science and Technology, Government of India. S.D. acknowledges the support from Ministry of Human Research and Development (MHRD), Govt. of India.

\end{document}